\documentstyle[epsfig]{mn}

\def\bi{\begin{itemize}}
\def\ei{\end{itemize}}
\def\be{\begin{equation}}
\def\ee{\end{equation}}
\def\ba{\begin{eqnarray}}
\def\ea{\end{eqnarray}}
\def\bse{\begin{subequations}}
\def\ese{\end{subequations}}
\def\M{{\cal {M}}}

\def\la{\langle}
\def\ra{\rangle}
\def\kB{k_{_B}}

\def\rhocr{\rho_{\textrm{\tiny{crit}}}}
\def\dd{\textrm{d}}

\title[Newtonian limit isothermal dark matter halos with
cosmological constant.]
{On the newtonian limit and cut--off scales of isothermal dark matter halos with
cosmological constant.}

\author[R. A. Sussman and X. Hernandez]
{Roberto A. Sussman$^1$ and Xavier Hernandez$^2$\\
$^1$ Instituto de Ciencias Nucleares,
Universidad Nacional Aut\'onoma de M\'exico
A. P. 70--543,  M\'exico 04510 D.F., M\'exico.\\
$^2$ Instituto de Astronom\'\i a,
Universidad Nacional Aut\'onoma de M\'exico
A. P. 70--264,  M\'exico 04510 D.F., M\'exico \\
}

\date{\today}

\begin{document}

\maketitle

\begin{abstract}

We examine isothermal dark matter halos in hydrostatic equilibrium with
a``$\Lambda$--field'', or  cosmological constant $\Lambda \, =
\,\Omega_\Lambda\,\rho_{\textrm{\tiny{crit}}}\,c^2$, where
$\Omega_\Lambda\simeq 0.7$,\, and  $\rho_{\textrm{\tiny{crit}}}$ is the present value of
the critical density
with $h\simeq 0.65$. Modeling cold dark matter as a self--gravitating Maxwell-Boltzmann gas, the
Newtonian limit of General Relativity yields equilibrium equations that are different from those
arising by merely coupling an ``isothermal sphere'' to the $\Lambda$--field within a Newtonian
framework. Using the conditions for the existence and stability of circular
geodesic orbits, the numerical solutions of the
equilibrium equations (Newtonian and Newtonian limit) show the existence of (I) an
``isothermal region'' ($0\leq r <r_2$), where circular orbits are stable and all variables
behave almost identically to those of an isothermal sphere; (II) an ``asymptotic region''
($r>r_1$)  dominated by the $\Lambda$--field, where the Newtonian potential oscillates and
circular orbits only exist in disconnected patches of the domain of $r$; (III) a
``transition region'' ($r_2\leq r <r_1$) between (I) and (II), where circular orbits exist
but are unstable. We also find that no stable configuration can exist with central density,
$\rho_c$, smaller than $2\,\Lambda$, hence any galactic haloes which virialized at $z< 30$
in a $\Lambda$--CDM cosmological background must have central densities of $\rho_c >
0.008\,\textrm{M}_\odot/\textrm{pc}^3$, in interesting agreement with rotation curve studies
of dwarf galaxies. Since
$r_2$ marks the largest radius of a stable circular orbit, it provides a characteristic
boundary or ``cut off'' maximal radius for isothermal spheres in equilibrium with a
$\Lambda$--field. For current estimates of $\rho_c$ and velocity
dispersion of virialized galactic structures, this cut off scale ranges from 90 kpc
for dwarf galaxies, up to to 3 Mpc for large galaxies and 22 Mpc for clusters. In a
purely  Newtonian framework these length scales are about 10\% smaller, though in either
case $r_2$ is between five and seven times larger than physical cut off scales
of isothermal halos, such as the virialization radius or the critical radius for the onset
of Antonov instability. These results indicate that the effects of the
$\Lambda$--field can be safely ignored in studies of virialized structures, but could be 
significant in the study of structure formation models and the dynamics of superclusters still in the linear regime or of
gravitational clustering at large scales ($r\approx 30$ Mpc).

\end{abstract}

\begin{keywords}
gravitation -- relativity -- galaxies: haloes -- galaxies: formation -- cosmology: theory
\end{keywords}

\section{INTRODUCTION}

Recent observational data from type Ia supernovae and CMB anisotropies suggests
that large scale cosmic dynamics are dominated by a repulsive type of matter--energy
density, generically known as a ``dark energy''. In its most simple
empirical formulation, dark energy can be described as a ``$\Lambda$--field'' associated
with the ``cosmological constant''. More sophisticated models of dark energy take the form
of scalar field sources in which the $\Lambda$--field corresponds to the
limit of vanishing scalar field gradients. This paper addresses an important issue in this
context, namely: to examine if this type of repulsive interaction has any influence on the
equilibrium of virialized galactic halo structures. Various aspects regarding this
question have been considered previously in the literature. Lahav
et al. (1991) have argued, using Newtonian spherical infall models (or
``top hat'' models), that the effects of a cosmological constant can be neglected in the
study of structure formation. Further improvements of ``top hat'' models incorporating
$\Lambda$ and/or quintessence--like fields have been considered
by Wang \& Steinhardt (1998), Iliev \& Shapiro (2001) and Lokas \& Hoffman (2001). Constraints on
$\Lambda$ at the level of clusters and superclusters have been placed by means of optical
gravitational lens surveys (Quast \& Helbig 1999) and peculiar velocities of galaxies
(Zehavi \& Dekel 1999).
Other papers (e.g. Axenides et al. 2000, Nowakowski 2001 and Nowakowski et al. 2001)
have studied the effect of $\Lambda$ on
galactic structures within the framework of General Relativity, though looking at this
effect not in the self--gravitating sources themselves, but through the dynamics of test
observers in a Schwarzschild-deSitter field that is supposed to be the ``external'' field
of these sources.

In this paper we complement and extend previous work by studying
a spherically symmetric, self--gravitating, Maxwell--Boltzmann (MB) gas in hydrostatic
equilibrium with a $\Lambda$--field given as a ``cosmological constant'',
subject to observational constraints. Although galactic halo structures are
characterized by Newtonian energies and velocities, we find it to be more correct to
examine these sources within General Relativity and then obtain the adequate (``weak
field'') Newtonian limit. The reason for this approach follows from the fact that in
Newtonian physics the mass density is the only source of  gravitation and so a
non--gravitational interaction, such as the
$\Lambda$--field, does not contribute to the ``effective gravitational mass''. In General
Relativity, on the other hand,  all forms of matter--energy  count as sources of
gravitation (curvature of spacetime). Therefore, even in the Newtonian limit, the
matter--energy contribution from the
$\Lambda$--field could be comparable (or at least might not be negligible in comparison)
to rest--mass density of the MB gas, so that the hydrostatic equations furnished by the
Newtonian limit might not coincide with those ``naively'' obtained from a Newtonian
framework. As we show in this paper, for a MB gas and a constant $\Lambda$--field, the
Newtonian limit yields a different set of equilibrium equations from those of Newtonian
theory. However, we cannot know {\it a priori} if this difference leads to significant or
negligible effects until we solve the equilibrium equations and compare their solutions in
both cases. Such ambiguity regarding the Newtonian limit would not arise if we were only
dealing with a MB gas (a thermal source), this is so because in this case the internal
energy (the only contribution to matter--energy besides rest--mass) is negligible in
comparison with rest--mass, leading to the ``isothermal sphere'' as the unequivocal
Newtonian limit.

It is a well known result that within the hierarchical model of structure formation (White \& Rees
1978) the final halo structure which results from the repeated merger events building up a
galaxy is well represented not by an isothermal structure, but by a centrally divergent
steep inner halo profile, first noted by Navarro, Frenk \& White (1997), and since confirmed
by numerous authors. However, controversy still surrounds this result in as much as direct dynamical
studies of galactic rotation curves, across wide dynamical ranges in galaxy size, appear to indicate
the presence of a constant density core in the dark haloes of galaxies (e.g. Burkert \& Silk 1997,
de Blok, McGaugh \& Rubin 2001, Firmani et al. 2001, among others). This being the case, we think
it worthwhile
to study in depth the modifications a simple isothermal halo suffers in the presence of a dominant
$\lambda$ component, even though this results may prove to be merely indicative of a case where not
just a stable halo is being treated, but including the full merger-driven formation of such a structure.
Whatever the correct structure scenario eventually turns out to be, a detailed understanding of
the behaviour of the quintessential galactic halo toy model -the isothermal halo- in the presence
of an important cosmological $\lambda$ term, will surely prove valuable.

The organization of the paper is described below. Sections II, III and IV
respectively provide the general relativistic formulation for a non--relativistic MB gas,
a $\Lambda$--field as a zero gradient limit of a quintessence--like scalar field and the
equations of hydrostatic equilibrium for these sources. We
examine the Newtonian limit of the models in section V,
showing that in the equilibrium equations in this limit do
not coincide with the corresponding Newtonian equations
for the same sources. In section VI we discuss the cut
off length scales, $r_1,\,r_2$, introduced by the $\Lambda$--field, comparing them
in section VII (by means of rough qualitative arguments) with physically motivated
length scales of virialized halo structures (virialization radius and critical radius
associated with Antonov instability). In section VIII we review briefly the equilibrium
equations for the King halos, while in section IX we use the numerical
integration of the various sets of equilibrium equations in order to verify the issues
discussed analytically and qualitatively in previous sections: the difference
between the Newtonian limit of General Relativity and a pure Newtonian framework and
the fact that  $r_2$ is 5--7 times larger than
physically motivated length scales. We briefly summarize these results in
section X. The paper contains 9 figures obtained from the numerical solution of
the equilibrium equations.

\section{The non--relativistic Maxwell-Boltzmann gas}

A relativistic MB gas follows from a Kinetic Theory formalism
associated with the equilibrium MB distribution (de Groot et al. 1980). For gas
particles of mass $m$, the parameter of ``relativity coldness'' is defined
as
\ba \beta \ \equiv \ \frac{m\,c^2}{\kB T},\label{betadef}\ea
where $\kB$ is Boltzmann constant and $T$ is the temperature. The
non--relativistic MB gas corresponds to the limit $\beta\gg 1$, leading to the
following forms for the energy density $\varepsilon$ and pressure $p$
\label{epspMB}\ba \varepsilon && = \ n\,\left(m\,c^2\,+
\frac{3}{2}\,\kB\,T\right)
\ = \ m\,c^2\,n\,\left(1+\frac{3}{2\beta}\right),\label{epsMB}\\
p && = \ n\,\kB\,T \ = \ \frac{m\,c^2\,n}{\beta},\label{pMB}\ea
with the particle number density, $n$, given by
\ba n \ = \
\left[\frac{m\,c}{\sqrt{2\,\pi\,\beta}\,\,\hbar}\right]^3\,
\exp\left[\beta\left(\frac{\mu}{m\,c^2}-1\right)\right],\label{nMB}\ea
where $2\pi\hbar$ is Planck's constant and $\mu$ is the Gibbs function per
particle (de Groot et al. 1980), the chemical potential. The MB distribution is a solution
of the Einstein--Boltzmann equations, irrespective of whether the gas is collisional or
collisionless, its applicability is only restricted by the criterion of ``non-degeneracy''
or ``dilute occupancy'' (Pathria 1972): \,\,  $\exp(\mu\beta/mc^2)\ll 1$. For typical
galactic halo variables, this criterion holds for particle masses complying with $m\gg 60$
eV (Cabral-Rossetti et al. 2002).

Under General Relativity, a spacetime whose source is a self gravitating MB gas
can be described by the Momentum--Energy tensor of a perfect fluid
\ba T_{\textrm{\tiny{MB}}}^{ab} \ = \ \varepsilon\,\,u^a\,u^b +
p\,h^{ab},\label{TMB}\ea
where $\varepsilon,\,p$ follows from equations (\ref{epsMB})--(\ref{nMB}) and
$h^{ab}=u^a\,u^b+g^{ab}$. Such a spacetime  must
admit a timelike Killing vector field (Maartens 1996)
\begin{equation}\beta^a = \beta\,u^a,\label{stat}\end{equation}
so that
\begin{equation}\dot
u_a-h_a^b\,(\ln\,\beta)_{,b} \ = \ 0,\label{Tolman}\end{equation}
where $\dot u_a=u_{;b}\,u^b$ is the 4-acceleration. Condition (\ref{stat})
implies that the spacetime must be stationary (or static if vorticity
vanishes), while condition (\ref{Tolman}) is the ``Tolman Law'' indicating that
equilibrium of a MB gas in General Relativity requires the existence of a nonzero
temperature gradient, a specific spacelike gradient that balances the
4-acceleration associated with a stationary frame. In a comoving representation,
condition (\ref{Tolman}) becomes $T\sqrt{-g_{tt}}=\textrm{const.}$

\section{Quintessence and the $\Lambda$ field}

``Dark energy'' matter-energy sources are usually represented as a quintessence--like
scalar field $\phi$ with a Momentum--Energy tensor
\ba T_{\textrm{\tiny{Q}}}^{ab} \ = \ \phi^{,a}\,\phi^{,b} \ - \
\left[\frac{1}{2}\,\phi_{,c}\,\phi^{,c}+V(\phi)\right]\,g^{ab},\label{TQ}\ea
where $V(\phi)$ is the scalar field potential. Since we are not interested in
a cosmological quintessence source, but rather in its gravitational interaction with
a MB gas at the galactic and galactic cluster scale, it is reasonable to assume that
the gradients $\phi_{,a}$ are negligible, hence we can approximate equation (\ref{TQ}) by
a ``cosmological constant'' or a $\Lambda$--field given by
\ba T_{\textrm{\tiny{Q}}}^{ab} \ \approx  \ -V(\phi_0)\,g^{ab} \ = \
-\Lambda\,g^{ab}, \label{TQL}\ea
equivalent to a source with matter--energy density
\ba \Lambda && = \ \Omega_\Lambda\,\rhocr\,c^2, \nonumber\\
\rhocr && = \ 1.88 \times
10^{-29}\,h^2\,\textrm{gm}/\textrm{cm}^3 \label{Lambda}
\ea
where, according to observational evidence, we can take $\Omega_\Lambda\simeq
0.7$ and $h\simeq 0.65$. If there is only gravitational
interaction between the MB gas and the $\Lambda$ field, the conditions
\ba T_{\textrm{\tiny{Q}}}^{ab}\,_{;b} \ = \ 0,\qquad
T_{\textrm{\tiny{MB}}}^{ab}\,_{;b} \ = \ 0,\label{Tcons}\ea
must hold separately. This condition is trivially satisfied for
$T_{\textrm{\tiny{Q}}}^{ab}$ given by equation (\ref{TQL}).

\section{The field equations}
In order to model an equilibrium galactic halo structure we will
consider spherically symmetric static spacetimes, compatible with equations (\ref{stat}) and
(\ref{Tolman}). The metric is then
\ba \dd s^2 \ =&&  -\,
\textrm{exp}\left(\frac{2\Phi}{c^2}\right)\,c^2\dd t^2+\frac{\dd r^2}{1-\,\kappa \,M/r}+
r^2\,\dd\,\Omega^2\nonumber\\
d\,\Omega^2 \ \equiv&& \ \dd\,\theta^2+\sin^2\theta\,\dd\,\phi^2,\label{metric}\ea
where $\Phi=\Phi(r),\,M=M(r)$ and $\kappa\equiv 8\pi G/c^2$, so that $M$ is given in mass
units. The source of equation (\ref{metric}) is the perfect fluid momentum--energy tensor
\ba T^{ab} \ =&& \
T_{\textrm{\tiny{MB}}}^{ab}+T_{\textrm{\tiny{Q}}}^{ab} \ =
(\varepsilon+\Lambda)\,\,u^a\,u^b + (p-\Lambda)\,h^{ab},\nonumber\\u^a \ =&& \
\exp\,(-\Phi/c^2)\,\delta^a\,_t,\label{EMtens}\ea
with $\epsilon,\,p$ given by equations (\ref{epsMB})--(\ref{nMB}).  Einstein's field equations
together with the balance law $T_{\textrm{\tiny{MB}}}^{ab}\,_{;b}=0$ for equation (\ref{metric}),
equations (\ref{EMtens}) and (\ref{Tolman2}) now take the form
\ba M' \ &&= \ (\varepsilon + \Lambda)\, r^2/c^2,\label{feq1}\\
\frac{\Phi'}{c^2} \ && = \
\frac{\kappa}{2}\,\frac{M+(p-\Lambda)\,r^3/c^2}{r\,(r-\kappa\,M)},\label{feq2}\\
p' \ &&= \ -(\varepsilon+p)\,\frac{\Phi'}{c^2},\label{beq}\ea
where a prime denotes  derivative with respect to $r$. Since the 4-acceleration is given by
$\dot u_a = \Phi'/c^2\,\delta_a^r$, the constraint (\ref{Tolman}) takes the form
\ba  \frac{\Phi'}{c^2} \ = \ \frac{\beta'}{\beta}\quad \Rightarrow\quad
\beta \ = \ \beta_c\,\exp\left(\frac{\Phi-\Phi_c}{c^2}\right),\label{Tolman2}\ea
where \, $\beta_c\equiv\beta(0)$ \, and \, $\Phi_c\equiv\Phi(0)$. The
subscript $_c$ will denote henceforth evaluation along the symmetry center
$r=0$. Inserting equations (\ref{epsMB}), (\ref{pMB}) and (\ref{nMB}) into equation (\ref{beq}) and using
condition (\ref{Tolman2}) yields after some algebra
\ba \frac{\mu'}{\mu}  = \ -\frac{\beta'}{\beta} \quad \Rightarrow \quad
\frac{\beta\,\mu}{m\,c^2} \ = \ C_0 \ = \ \textrm{const.},\label{condmu_rel}\ea
so that $n,\,\epsilon$ and $p$ become functions depending only on $\beta$ (or on $\Phi$ via
equation (\ref{Tolman2})). Using equation (\ref{condmu_rel}) and re-scaling $n$ with
respect to the central rest--mass density $\rho_c = m\,n_c$, the matter--energy density
and pressure for the equilibrium MB gas become
\ba \varepsilon \ = \ \rho\,c^2\,
\left(1+\frac{3}{2\,\beta}\right),\qquad p \ = \ \frac{\rho\,c^2}{\beta},\label{epMB}\ea
where the rest--mass density is given by
\ba
\frac{\rho}{\rho_c} \ =
\left(\frac{\beta_c}{\beta}\right)^{3/2}\,\textrm{e}\,^{\beta_c-\beta}.\label{rhoMB}\ea
However, it is more useful to express these state variables and the field equations
themselves in terms of the central velocity dispersion (e.g. Binney \& Tremaine 1987)
\ba
\sigma_c^2  \equiv \ \frac{\kB\,T_c}{m} \ = \
\frac{c^2}{\beta_c},\label{sigmac}\ea
and the dimensionless variable
\ba \Psi \ \equiv \ \frac{\Phi_c-\Phi}{\sigma_c^2},\label{defPsi}\ea
which in the Newtonian limit becomes the ``normalized potential''.
Combining equations (\ref{Tolman2}), (\ref{sigmac}) and (\ref{defPsi}), we obtain the
``redshifted'' temperature provided by Tolman's law
\ba T \ = \ T_c\,\exp\left(\frac{\Phi_c-\Phi}{c^2}\right) \ = \
T_c\,\exp\left(\frac{\sigma_c^2}{c^2}\,\Psi\right),\label{temp}\ea
showing the specific temperature gradient of an MB
gas in thermodynamic equilibrium.

With the help of equations (\ref{Tolman2}) and (\ref{rhoMB}), we obtain
\ba \frac{\rho}{\rho_c} \ =
\ \exp\left[\frac{3}{2}\,\frac{\sigma_c^2}{c^2}\,\,\Psi+\frac{c^2}{\sigma_c^2}
\left(1-\textrm{e}^{-(\sigma_c/c)^2\,\Psi}\right)\right],\nonumber\\
\label{eqn}\ea
so that $\varepsilon$ and $p$ become
\label{eqep}\ba
\varepsilon&& \ =  \
\rho\,c^2\,\left[1+\frac{3\,\sigma_c^2}{2\,c^2}\,\textrm{e}^{(\sigma_c/c)^2\,\Psi}\right],\label{eqe}
\\ p&& \ = \
\rho\,\sigma_c^2\,\textrm{e}^{-(\sigma_c/c)^2\,\Psi},\label{eqp}\ea
The field equation that we have to solve are then equations
(\ref{feq1}) and (\ref{feq2}), with $\varepsilon,\,p$ and $\Lambda$ given by equations (\ref{Lambda}),
(\ref{eqe}) and (\ref{eqp}).  However, it is convenient to express these equations in terms of equations
(\ref{sigmac}), (\ref{defPsi}) and the following dimensionless variables
\ba x&& \ = \ \frac{r}{r_0}, \qquad r_0^{-2} \ = \ \frac{4\pi
G\,\rho_c}{\sigma_c^2}, \nonumber\\
\M&& \ = \ \frac{M}{\rho_c\,r_0^3},\label{adimvars}
\ea
leading to
\ba \frac{\dd \M}{\dd x}&&  = \
\left[\frac{\rho}{\rho_c}\,\left(1+\frac{3\,\sigma_c^2}{2\,c^2}\,
\textrm{e}^{(\sigma_c/c)^2\Psi}\right)+\lambda\,\right]\,x^2,\nonumber\\
\frac{\dd \Psi}{\dd x}&&  = \
-\frac{\M+\left[(\rho/\rho_c)(\sigma_c^2/c^2)\,\textrm{e}^{(\sigma_c/c)^2\Psi}\,-\,
\lambda\,\right]\,x^3}{x\,\left[\,x-2\,(\sigma_c^2/c^2)\,\M\right]},\nonumber
\\\label{feqrel}\ea
where $\rho/\rho_c$ follows from equation (\ref{eqn}) and
\ba \lambda \ \equiv \ \Omega_\Lambda\,
\frac{\rhocr}{\rho_c} \ = \ 2.538 \, \times \, 10^{-7}\,
\Omega_\Lambda\,h^2\,\frac{\textrm{M}_\odot/\textrm{pc}^3}{\rho_c},\label{lambda}\ea
while the length scale $r_0$ is one third of the ``King radius'' of Binney \& Tremaine (1987)
and can be given numerically as
\ba r_0 \ = \ 4.2303 \times 10^{-6} \, \frac{\sigma_c}{\textrm{km/sec}}\,
\left(\frac{\textrm{M}_\odot/\textrm{pc}^3}{\rho_c}\right)^{1/2}\,\textrm{Mpc}.
\label{r0vals}\ea
Now equation (\ref{feqrel}) is a self--consistent system of differential equations that can be
integrated once numerical values for the independent  parameters
($\rho_c,\,\sigma_c,\,\Omega_\Lambda,\,h$) have been supplied, while regularity conditions
at the center require $\M_c=\M'_c=\Psi_c=\Psi\,'_c=0$. Notice that by setting
$\Omega_\Lambda=0$, we obtain the general relativistic equations of hydrostatic equilibrium
for a Maxwell--Boltzmann gas characterized by the equation of state of equations
(\ref{epsMB}) and (\ref{pMB})--(\ref{nMB}).

An alternative system to that of equation
(\ref{feqrel}), that is easier to handle numerically, follows by expressing
$\rho,\,\varepsilon,\,p$ as functions of $\beta$ by means of equations (\ref{epMB}) and
(\ref{rhoMB}), so that $\beta$ replaces $\Psi$ in equation (\ref{feqrel}). However, the variables
$\M$ and $\Psi$ in equation (\ref{feqrel}) are more convenient to explore the Newtonian
limit.

\section{The Newtonian limit vs. Newtonian configurations}

We can identify Newtonian conditions, within the framework of General Relativity, as the
weak field limit of the theory for the metric of equation (\ref{metric}) and source
of equation (\ref{EMtens}). By looking at equations (\ref{temp}), (\ref{eqn}), (\ref{eqe})-- (\ref{eqp}) and
(\ref{feqrel}), dimensionless variables $\Psi$ and $\M$ (which are
not, necessarily, small) always appear multiplied by the ratio
$\sigma_c^2/c^2$. Since $\sigma_c$ is a characteristic velocity of the MB gas, we
can examine Newtonian conditions by expanding all relevant quantities in powers of this
ratio. Using relation (\ref{adimvars}) we find to first order in
$\sigma_c^2/c^2$

\ba \exp\left[\frac{2\,(\Phi-\Phi_c)}{c^2}\right] \ \approx \
1-2\,\frac{\sigma_c^2}{c^2}\,\Psi,\nonumber\\
\left[1-\frac{\kappa\,M}{r}\right]^{-1}\ \approx \ 1+2\,
\frac{\sigma_c^2}{c^2}\,\frac{\M}{x}. \label{weakfield} \ea
Thus, the Newtonian limit of General Relativity for the sources under consideration is
characterized by
\ba \frac{\sigma_c}{c} \ \ll \ 1\qquad \Rightarrow\qquad \frac{\Phi}{c^2} \ \ll \ 1,\qquad
\frac{\kappa\,M}{r}\ \ll \ 1,\qquad \label{newtconds}\ea
so that $g_{tt}\approx-1,\,\,g_{rr}\approx 1$ and we can identify (in this limit) the
Newtonian potential with $\Phi$. Also, at first order in $\sigma_c^2/c^2$ we have
\ba &&-\frac{3}{2}\,\frac{\sigma_c^2}{c^2}\,\,\Psi+\frac{c^2}{\sigma_c^2}
\left[1-\textrm{e}^{-(\sigma_c/c)^2\,\Psi}\right] \approx \nonumber\\ &&\frac{c^2}
{\sigma_c^2} \left[1-\textrm{e}^{-(\sigma_c/c)^2\,\Psi}\right] \approx
\frac{c^2}{\sigma_c^2} \left[1-\left\{1-\frac{\sigma_c^2}{c^2}\,\Psi\right\}\right]
\approx \Psi,\nonumber\ea
and so equations (\ref{temp}), (\ref{eqn}) and (\ref{eqe})--(\ref{eqp}) become
\ba \frac{\rho}{\rho_c} \ \approx \
\textrm{e}^\Psi\label{rho_newt},\quad
\varepsilon \ \approx \ \rho\,c^2,\quad p \ \approx \  \frac{\sigma_c^2}{c^2}\,
\varepsilon \ \ll \ \varepsilon,\label{newtstv}\ea
\ba T \ \approx \
T_c\,\left[1-\frac{\sigma_c^2}{c^2}\,\Psi\right]
\ = \
T_c\,\left[1-\frac{\sigma_c^2}{c^2}\,\ln\,\frac{\rho}{\rho_c}\right]\label{tempNL}\ea
thus, a spherical MB gas in its Newtonian limit approaches the
``isothermal sphere'' with $T=T_c$ (or, equivalently $\sigma=\sigma_c$) for all $x$ (see
figure 1). The field equations (\ref{feqrel}) become (at first order in $\sigma_c^2/c^2$)
the system
\label{feqnewt1}
\ba \frac{\dd \M}{\dd x}&&  = \
\left[\textrm{e}^{\Psi}
\left(1+\frac{3\,\sigma_c^2}{2\,c^2}\right)+\lambda\,\right]\,x^2,\nonumber\\
\frac{\dd \Psi}{\dd x}&&  = \
-\frac{\M}{x^2}-\left(\frac{\sigma_c^2}
{c^2}\,\textrm{e}^{\Psi}-\lambda\,\right)\,x,\label{feqnewt1}\ea
where we have kept, together with rest--mass energy density ($\propto\rho$), the other two
contributions to the total matter--energy density: internal energy ($\propto
p\propto \sigma_c^2\,\rho$) and that of the $\Lambda$ field ($\lambda$), as we do not
know \textit{a priori} which one of the latter two contributions is larger. In the case
without the $\Lambda$ field ($\lambda=0$), we only need to compare rest--mass and
internal energy densities. From equations (\ref{newtconds}) and (\ref{newtstv}), it is evident
that in this case the Newtonian limit of equation (\ref{feqnewt1}) can only lead to the
equilibrium equations for the Newtonian ``isothermal sphere''
\ba \frac{\dd \M}{\dd x}&& \ = \
\textrm{e}^{\Psi}\,x^2,\nonumber\\
\frac{\dd \Psi}{\dd x}&& \ = \
-\frac{\M}{x^2},\label{feqnewt2}\ea
whose corresponding Poisson equation is
\ba -\frac{1}{x^2}\,\frac{\dd}{\dd x}\left(x^2\,\frac{\dd \Psi}{\dd x}\right) \ = \
\textrm{e}^\Psi.\label{poissonIS}\ea
However, if we have the contribution of the $\Lambda$ field
(besides the internal--energy term) then, before deciding which terms can be neglected,
we should verify first if $\lambda$ is comparable in magnitude to the internal energy
density term that is first order in $\sigma_c^2/c^2$. This comparison is especially
important in the ``force'' equation ({\it i.e.} the second of equations (\ref{feqnewt1})) 
where this term and $\lambda$ have opposite signs.

For a wide variety of galactic objects (from dwarf
galaxies to clusters), we have the range $4\,\textrm{km/sec} < \sigma_c <
1500\, \textrm{km/sec}$ (e.g. Kleyna 2002, Wilkinson 2002, Salucci \& Burkert 2000), hence
\ba 10^{-9} \ < \ \frac{\sigma_c^2}{c^2} \ < \ 2.5\times
10^{-5},\label{sigmavals}\ea
On the other hand, for currently accepted estimates of central
halo density: $0.001\,\,M_\odot/\textrm{pc}^3
< \rho_c < 1\,\,M_\odot/\textrm{pc}^3 $, \,\, (e.g. Firmani et al. 2001, Shapiro \& Iliev 2002,
Dalcanton \& Hogan 2001)
we have
\ba 4\times 10^{-6}\,\Omega_\Lambda h^2 \ < \ \lambda \ < \ 4\times
10^{-3}\,\Omega_\Lambda h^2.\label{xivals}\ea
Therefore, since $\Omega_\Lambda\simeq 0.7$ and $h\simeq 0.65$, it is evident that near the
center we have $\rho/\rho_c=\textrm{e}^\Psi \approx 1$ and so $\sigma_c^2/c^2$ and
$\lambda$ can have comparable magnitude (especially for structures with large $\rho_c\sim
1\,\textrm{M}_\odot /\textrm{pc}^3$).  However, as
$x$ increases the $\lambda$ term remains constant while $\rho/\rhocr$ decreases, implying
that for large $x$ the constant $\lambda$ term will always end up dominating. Also, since
the Newtonian potential is negative with $|\Phi_c|>|\Phi|$ so that $\Psi<0$, we can safely
assume that $\lambda\gg (\sigma_c/c)^2\,\textrm{e}^\Psi$ holds even for small values of $x$.

Therefore, we can ignore the contribution from hydrostatic pressure in equations
(\ref{feqnewt1}) and replace them by
\ba \frac{\dd \M}{\dd x}&& \ = \
\left(\,\textrm{e}^{\Psi}+\lambda\,\right)\,x^2,\nonumber\\
\frac{\dd \Psi}{\dd x}&& \ = \
-\frac{\M}{x^2}+\lambda\,x,\label{feqnewt3}\ea
so that
\ba -\frac{1}{x^2}\,\frac{\dd}{\dd x}\left(x^2\,\frac{\dd \Psi}{\dd x}\right) \ = \
\textrm{e}^\Psi-2\,\lambda.\label{poissonNL}\ea
These equations are the Newtonian limit of equation (\ref{feqrel}), based on removing
only those terms that can be neglected for all $x$ in terms of expansions around
$\sigma_c^2/c^2$ and of comparison with rest--mass density. However, it is interesting to
remark that if we start from the Newtonian isothermal sphere, as given by
equations (\ref{feqnewt2}) and (\ref{feqnewt2}), and within a purely Newtonian framework
(\textit{ie} not the Newtonian limit of General Relativity) we generalize it by adding a
$\Lambda$-type repulsive interaction, we would not obtain equations (\ref{feqnewt3}) and
(\ref{poissonNL}), but the set

\

\ba \frac{\dd \M}{\dd x}&& \ = \
\textrm{e}^{\Psi}\,x^2,\nonumber\\
\frac{\dd \Psi}{\dd x}&& \ = \
-\frac{\M}{x^2}+\lambda\,x,\label{feqnewt4}\ea
\ba -\frac{1}{x^2}\,\frac{\dd}{\dd x}\left(x^2\,\frac{\dd \Psi}{\dd x}\right) \ = \
\textrm{e}^\Psi-3\,\lambda,\label{poissonPN}\ea
since in Newtonian gravity the $\Lambda$ field only enters into the
``force'' equation in (\ref{feqnewt4}) but does not enter in the evaluation of the
gravitational mass $\M$ that follows from integrating equation (\ref{feqnewt4}).
On the other hand, in the Newtonian limit equations
(\ref{feqnewt3}) the form of $\M$ (and thus the forms of $\Psi$ and all variables related
to $\M$ and $\Psi$) is affected by the presence of the $\lambda$ term.

This difference between a purely Newtonian treatment and the Newtonian limit from General
Relativity is a significant feature certainly worth remarking, since it concerns $M$ which
is the effective gravitational mass affecting the evolution of test observers and light
rays. Since General Relativity is currently the best available theory of gravitation, we
suggest that the system of equations (\ref{feqnewt3})--(\ref{poissonNL}) should be preferred (at least
conceptually) to the purely Newtonian system of equations (\ref{feqnewt4})--(\ref{poissonPN}). In fact,
these systems have different solutions and if the differences in these solutions are not
negligible, this can (in principle) be relevant in dark matter tracing and gravitational
lensing methods used for estimating the properties of dark matter haloes of galactic
clusters. The long numerical integrations of cosmological N-body simulations imply that even
minor correction factors, of the type discussed here, could have important effects on the final
result. Figures 2a and 2b illustrate how the correct Newtonian limit of a
Maxwell--Boltzmann gas in equilibrium with a $\Lambda$--field is furnished by equations
(\ref{feqnewt3})--(\ref{poissonNL}), just as the Newtonian limit without $\Lambda$ is the
isothermal sphere. However, before solving equations (\ref{feqnewt3}) and (\ref{feqnewt4}) numerically
it is not possible to know if the differences in their solutions are significant or
negligible. We will examine this issue in section IX (see figures 2 and 3). For the remaining of the
discussion we will consider only the Newtonian limit (NL) equations (\ref{feqnewt3}).

\section{Existence and stability of circular geodesic orbits}

The relativistic generalization of the ``rotation velocity'' along circular orbits is the
rotation velocity of test observers along circular geodesics in the spacetime, equation
(\ref{metric}). Bounded timelike geodesics of equation (\ref{metric}) are characterized
by (Chandrasekhar 1983) \,\, $\theta=\pi/2$ and by two constants of motion, ``energy'' $E_0 =
2\,\textrm{e}^{2\,\Phi/c^2}\,\dot t$ and ``angular momentum'' $L_0  =  r^2\,\dot\varphi$,
where the dot denotes derivative with respect to proper time of the observers. Radial
motion is governed by
\ba \frac{\textrm{e}^{2\,\Phi/c^2}}{1-\kappa\,M/r}\,\dot r^2+V(r) \ = \
E_0^2,\nonumber\ea
where
\ba V(r) \ = \
\textrm{e}^{2\,\Phi/c^2}\left(1+\frac{L_0^2}{r^2}\right),\nonumber\ea
is the ``effective potential''. The conditions for the existence of circular
geodesic orbits with radius $r=r^*$ are $\dot r = \ddot r = 0$, leading to
\ba E_0^2 \ = \ \frac{\textrm{e}^{2\,\Phi/c^2}}{1-r\,\Phi'/c^2},\qquad L_0^2 \ = \
\frac{r^3\,\Phi'/c^2}{1-r\,\Phi'/c^2},\label{EL_forms}\ea
so that
\ba \Phi' \ > \ 0, \qquad 1-r\,\Phi'/c^2 \ > \ 0,\label{conds_1}\ea
Stability of these orbits requires $r=r^*$ to be a minimum of the effective
potential. Hence we must have $V'(r^*)=0$ and $V''(r^*)>0$, leading (with the help of
equations (\ref{EL_forms})) to
\ba 3\,\Phi' + r\,\Phi'' - 2r\,\Phi'{}^2/c^2 \ > \ 0.\label{conds_2}\ea
Using the full relativistic equilibrium equations (\ref{feqrel}) and expressing
conditions (\ref{conds_1}) in terms of the variables introduced in section IV, the conditions
for the existence of circular geodesic orbits are, up to first order in $\sigma_c^2/c^2$,
\ba \frac{\lambda\,x^3-\M}{x^2}
-\left[x\,\textrm{e}^\Psi+\frac{2\,\M}{x^3}-2\,\lambda\,\M\right]\,\frac{\sigma_c^2}{c^2}
\ < \ 0,\label{cond_11}\\ 1+\frac{\lambda\,x^3-\M}{x}\,\frac{\sigma_c^2}{c^2} \ > \
0,\label{cond_22}\ea
while the condition (\ref{conds_2}) for stability becomes
$${{(3\lambda -\textrm{e}^\Psi )\,x^3-M} \over {x^2}}+{{\sigma _c^2}
\over {c^2}}
\left\{ {{{} \over {}}4\lambda ^2\,x^3+(x^3\,\textrm{e}^\Psi
-2\M)\,\lambda } \right.$$
\ba\left. {+\left[ {{{(\Psi ^2-3\Psi -11)\,x-6\M} \over 2}-3\M}
\right]\,\textrm{e}^\Psi
+{{2\M} \over {x^3}}} \right\}<0,\nonumber\\\label{cond_33}\ea
In the Newtonian limit we can neglect the terms multiplying $\sigma_c^2/c^2$, hence
condition (\ref{cond_22}) becomes trivial, while conditions (\ref{cond_11}) and
(\ref{cond_33}) respectively become the Newtonian limit conditions for existence and
stability of circular geodesic orbits
\ba \M - \lambda\,x^3 \ > \ 0,\label{condexist}\\
\M -(3\,\lambda-\textrm{e}^\Psi)\,x^3 \ > \ 0.\label{condstab}\ea
These conditions coincide with those that would have resulted had we used the Newtonian
limit equations (\ref{feqnewt3}) (instead of the relativistic equations) in equations
(\ref{conds_1}) and (\ref{conds_2}). If we had used the Newtonian equations
(\ref{feqnewt4}) we would have obtained the same existence condition (\ref{condexist})
but the stability condition would be condition (\ref{condstab}), modified by having the $\lambda$
term multiplied
by a factor of $4$, instead of $3$.  Figures 3a and 3b depict graphically conditions
(\ref{condexist}) and (\ref{condstab}) from the numerical solution of conditions (\ref{feqnewt3}) and
(\ref{feqnewt4}).

The actual velocity of observers along circular geodesics follows from evaluating $
\dd \varphi/\dd t= \dot\varphi/\dot t$, which with the help of the previous equations
(see Cabral-Rosetti et al. 2002) yields
\ba \textrm{v}^2(r) \ = \ r\,\Phi' \ = \
-x\,\sigma_c^2\,\frac{\dd\,\Psi}{\dd\,x} \ = \
\sigma_c^2\,\left[\frac{\M}{x}-\lambda\,x^2\right]\label{rotvel}\ea
where we have used equation (\ref{feqnewt3}) to eliminate $\dd \Psi/\dd x $.
Since $\Phi$ becomes the Newtonian gravitational potential in the Newtonian limit,
circular geodesics orbits exist and $\textrm{v}(r)$ is well defined as long as the
gravitational force field is attractive (\textit{ie} $\Phi' > 0$ or equivalently
$\dd\Psi/\dd x< 0$).

For physically reasonable spherical configurations, we must demand that existence and
stability conditions (\ref{condexist}) and (\ref{condstab}) (or (\ref{condexist_ave}) and
(\ref{condstab_ave})), as well as that $\textrm{v}^2>0$,  should hold at least in a finite
range $0<x< x_\lambda$ around the symmetry center, where $x_\lambda$ is a radius where one
of these conditions breaks down. A sufficient condition for this follows by looking at the
behavior of conditions (\ref{condexist}) and (\ref{condstab}) around the symmetry center. Expanding
$\M$ around $x=0$ with the help of equation (\ref{feqnewt3}) and the regularity conditions
$\M(0)=\M_{x}(0)=\Psi(0)=\Psi_x(0)=0$, where the subindex
$_x$ denotes derivative with respect to $x$. We obtain $\M \,\approx
(1/6)\M_{xxx}(0)\,x^3 \,=\, (1+\lambda)\,x^3/3$, hence for
$x\approx 0$ both conditions (\ref{condexist}) and (\ref{condstab}) yield
$(1-2\,\lambda)\,x^3/3 > 0$, so that the central density must satisfy
\ba \rho_c \ > \
2\,\Omega_\Lambda\,\rhocr.\label{min_lam}\ea
Had we used the Newtonian equations (\ref{feqnewt4})
instead of equation (\ref{feqnewt3}), we would have obtained $\rho_c> 3\,\Omega_\Lambda\,\rhocr$,
representing a very small departure from the Newtonian limit expression  (\ref{min_lam}).

The fact that physically reasonable configurations require a minimal bound on $\rho_c$ is
an interesting result that has no equivalent in the isothermal sphere and the King halos.
Since $\textrm{e}^\Psi=\rho/\rho_c$ and setting
$\lambda=0$ in equations (\ref{condexist}), (\ref{condstab}) and (\ref{rotvel}), it is evident that
stable geodesic circular orbits, of any radius and for any value of $\rho_c>0
$, always exist for the isothermal sphere and the King models (see section VIII) as long
as $\M$ and $\rho$ are positive. If there is a $\Lambda$--field then for very small
central densities violating condition (\ref{min_lam}), conditions (\ref{condexist}) and
(\ref{condstab}) do not hold in a range around the symmetry center, though these
conditions do hold for a set of ranges like $0<x_i<x<x_j$ (see figure 3c). However, these
cases with $\rho_c\leq 2\,\Omega_\Lambda\,\rhocr$ are unphysical and
will not be examined any further.

Bearing in mind equations (\ref{Lambda}) and (\ref{lambda}), condition (\ref{min_lam}) is satisfied
by all known virialized galactic structures whose current estimates of $\rho_c$ are
between three and six orders of magnitude larger than $\rhocr$. In fact, the extremely low
values $\rho_c<2\Omega_\Lambda\,\rho_{\textrm{\tiny{crit}}}$ would apply today only to
large superclusters, still ``in--falling'' within a linear regime, whose density contrast
is of order unity. Obviously such large scale structures
cannot be  equilibrium configurations (hence the remarks in Nowakowski 2001, Nowakowski
et al. 2002 are
obvious for present day structures).  Moreover, $\rho_{\textrm{\tiny{crit}}}$ is a (model
dependent) function of the cosmic era, hence the minimal $\rho_c$ for stable
configurations varies with the virialization redshift
$z_{\tiny{\textrm{vir}}}$. Assuming a $\Lambda$--CDM model for which
$\Omega_{\tiny{\textrm{CDM}}}+\Omega_\Lambda=1$, the critical density at any $z$ is given in terms
of $\rho_{\textrm{\tiny{crit}}}$ today by
\ba \rho_{\textrm{\tiny{crit}}}(z) \ = \
\left[(1-\Omega_\Lambda)\,(1+z)^3+\Omega_\Lambda\right]\,\rho_{\textrm{\tiny{crit}}}(z=0).
\label{rhocrz}\ea
Thus, for dwarf galaxies which virialized at large $z$ (\textit{i.e.} $z_{\tiny{\textrm{vir}}} <
30$), equations (\ref{min_lam}) and (\ref{rhocrz}) imply
\ba \rho_c(z=30) \ > \ 1.8\,\times\,10^4\,\rho_{\textrm{\tiny{crit}}}(z=0) \ \sim \
0.008\,\textrm{M}_\odot/\textrm{pc}^3,\nonumber\\\label{minrhoc}\ea
a minimal bound on $\rho_c$ that agrees with current estimates (Firmani et al. 2001,
Shapiro \& Iliev 2002, Kleyna et al. 2002) for central halo
densities of these galaxies. Within the current hierarchical models of structure formation, where
large structures form as the result of the merger of smaller ones, given Liouville's theorem, it is
clear that the central densities of the first bound structures will also represent an indicative
upper bound on the central densities of larger structures. It is interesting that dynamical
studies across wide ranges of galactic haloes consistently find central densities above the
stability requirement at $z \approx 30$ of $0.008 M_{\odot} pc^{-2}$.
Figure 4 illustrates the minimal bound on $\rho_c$ given by condition
(\ref{minrhoc}).

\section{Cut off scale of $\Lambda$ and comparison with other length scales}

Assuming that condition (\ref{min_lam}) holds, each one of conditions (\ref{condexist}) and
(\ref{condstab}) will necessary break down at some radii, $x_1,\, x_2$,
given by the smallest positive roots of the equations
\ba \M(x_1) - \lambda\,x_1^3 \ = \ 0,\label{xlambda_1}\\
\M(x_2) -[3\,\lambda-\textrm{e}^{\Psi((x_2)}]\,x_2^3 \ = \
0.\label{xlambda_2}\ea
These radii provide ``cut off'' length scales that do not exist when
$\lambda=0 $. We will compare these length scales with other physical cut off scales,
such as the virialization radius and the radius for the onset of Antonov instability.

It is useful to express conditions (\ref{condexist}) and (\ref{condstab}) in terms of the
volume average of rest--mass energy density $\rho$. Under Newtonian conditions the total
matter--energy density is $\varepsilon_{\textrm{\tiny{total}}}
=(\rho+\Omega_\Lambda\rhocr)\,c^2 $, while its volume average is
\ba  \frac{\la\varepsilon_{\textrm{\tiny{total}}}\ra}{c^2}&& \ = \
\frac{4\,\pi\,\int_0^r{(\rho+\Omega_\Lambda\,\rhocr)\,r^2\dd
r}}{4\,\pi\,\int_0^r{r^2\dd r}} \ = \ \frac{3\,M}{r^3}\nonumber\\
&&= \ \la\rho\ra + \Omega_\Lambda\rhocr,\label{rhoav}\ea
therefore, conditions (\ref{condexist}) and
(\ref{condstab}) can be given respectively as
\ba \la \rho \ra && > \ 2\,\Omega_\Lambda\,\rhocr,\label{condexist_ave}\\
\frac{1}{4}\,\la\rho\ra + \frac{3}{4}\,\rho&& > \
2\,\Omega_\Lambda\,\rhocr,\label{condstab_ave}\ea
where we used equation (\ref{lambda}) and $M/r^3=\rho_c\,\M/x^3$. Equations (\ref{xlambda_1}) and
(\ref{xlambda_2}) then become 
\ba \la \rho \ra (x_1) && = \ 2\,\Omega_\Lambda\,\rhocr,\label{xlambda1_ave}\\
\frac{1}{4}\,\la\rho\ra(x_2) + \frac{3}{4}\,\rho(x_2)&& = \
2\,\Omega_\Lambda\,\rhocr,\label{xlambda2_ave}\ea
From the volume averaging definition (\ref{rhoav}), we obtain by integration by parts
\ba \la\rho\ra \ = \ \frac{\int{\rho\,x^2\,\dd x}}{\int{x^2\,\dd x}} \ = \
\rho-\frac{\int{x^3\,\rho_x\,\dd x}}{x^3}\ea
Hence, for $0<x< x_\lambda$, where $x_\lambda=r_0\,x_\lambda$ is either one of
$x_1,\,x_2$, we have have: $\la\rho\ra\geq \rho$. Since condition
(\ref{condexist_ave}) holds in this range, then $\rho_x<0$ follows from
$\Psi_x<0$ (because $\rho_x\propto\Psi_x\,\exp(\Psi)$). Therefore for any $x$ in the
range of interest we have $(1/4)\,\la\rho\ra +(3/4)\,\rho\geq \la\rho\ra$. Hence, the
root $x_2$ corresponds to larger density values than
$x_1$, but since both $\rho$ and $\la\rho\ra$ decrease with $x$, we must
necessarily have
\ba x_2 \ \leq \ x_1,\label{xlambda12}\ea
which is a reasonable result, since it is to be expected that the repulsive
$\Lambda$--field will destabilize those circular orbits with radii $x_2 < x<
x_1$ that lie near the length scale $x_1$ beyond which no such
orbits exist. As we show in section IX, the range $x_2 < x< x_1$
corresponds to a transition zone between the isothermal region  $0<x<x_2$ and
the asymptotic region $x> x_1$ dominated by the $\Lambda$--field. Thus, since $x=x_2$
marks the maximal radius of a stable circular orbit, we can consider
this radius as a ``cut off'' scale for isothermal spheres in hydrostatic equilibrium with
a $\Lambda$--field. In the remaining of this section we will use simple qualitative
arguments to compare other cut off length scales of the isothermal sphere with
$x_1$. Since numerical estimates (see figure 5a) show that $x_2$ is about two thirds of
$x_1$ (see section IX), this comparison can be extended to $x_2$ as well.

Consider the virialization radius, $r_{\textrm{\tiny{vir}}}$, the currently accepted
physical cut off scale associated with the virialization of galactic
structures (Padmanabhan 1995). The condition to determine the value of  this length scale
follows from the Newtonian spherical ``in fall'' (or ``top hat'') model
as Lokas \& Hoffman (2001), Iliev \& Shapiro (2001) and Padnamabhan (1995) show,
\ba \la\rho\ra (r_{\textrm{\tiny{vir}}})|_{z_{\textrm{\tiny{vir}}}} \ \approx \
\Delta\,\rhocr(z_{\textrm{\tiny{vir}}}),\label{rvir}\ea
or, equivalently (using equations (\ref{adimvars}) and (\ref{rhoav})), as
\ba
\frac{\M(x_{\textrm{\tiny{vir}}})|_{z_{\textrm{\tiny{vir}}}}}{x_{\textrm{\tiny{vir}}}^3}
\
\approx \ \frac{\Delta
+\Omega_\Lambda}{3}\,\frac{\rhocr(z_{\textrm{\tiny{vir}}})}{\rho_c(z_{\textrm{\tiny{vir}}})},
\label{rvir2}\ea
where the factor $\Delta$ is a model dependent (see Lokas \& Hoffman 2001 for a
discussion on this issue)
contrast factor between the average rest--mass density of the overdense region and the critical
density of the cosmological background, both evaluated at a virialization epoch,
$0<z_{\textrm{\tiny{vir}}} < 30$, which depends on the galactic structure and/or a
specific structure formation model. From equations (\ref{xlambda1_ave}) and
(\ref{rvir}), we can see that both
$\la\rho\ra(r_1)$ and $\la\rho\ra(r_{\textrm{\tiny{vir}}})$ are proportional to
$\rhocr$. Therefore, for all $z<z_{\textrm{\tiny{vir}}}$, once the galactic structure
has virialized and the equilibrium equations are valid, we have
\ba \frac{\la\rho\ra(r_1)}{\la\rho\ra(r_{\textrm{\tiny{vir}}})} \ \approx \
\frac{2\,\Omega_\Lambda}{\Delta},\label{rvir0}
\ea
or the same expression with a factor 3 instead of 2 if we had used the
Newtonian equations (\ref{feqnewt4}).  Since  it is reasonable to assume (to a good
approximation) that $\la\rho\ra\sim\rho\sim r^{-2}$ for $r<r_1$, a rough
qualitative estimate of the ratio $r_1/r_{\textrm{\tiny{vir}}}$ is
\ba \frac{r_1}{r_{\textrm{\tiny{vir}}}} \ \sim \
\left[\frac{\Delta}{2\,\Omega_\Lambda}\right]^{1/2},
\label{rlv}\ea
Considering that $\Delta \sim 100$ in the case when the cosmological
background complies with $\Omega_{\textrm{\tiny{CDM}}}+\Omega_\Lambda=1$ (Lokas \& Hoffman 2001),
this estimate means that for present day virialized structures $r_1$ is
approximately one order of magnitude larger than
$r_{\textrm{\tiny{vir}}}$. Because of equation (\ref{xlambda12}), $r_2$ should be less than an order
of magnitude larger than $r_{\textrm{\tiny{vir}}}$. The precise numerical value of
$x_{\textrm{\tiny{vir}}}$ is displayed in figure 5b, while the qualitative estimates
mentioned before are corroborated numerically in figures 6 and 7, showing that
$r_2$ is about seven times larger than $r_{\textrm{\tiny{vir}}}$. Using the
Newtonian equations (\ref{feqnewt4}) yields smaller $r_1$ and $r_2$ by a factor of
$\sqrt{2/3}$. In either case, these results indicate that the $\Lambda$--field should have a
negligible effect on the virialization process.

Another physically motivated length scale, characteristic of the isothermal
sphere, is the critical radius associated with the Antonov instability, or with the onset
of the gravothermal catastrophe. This radius follows from demanding that the
structural parameters of an isothermal sphere should correspond to a stable thermal
equilibrium, characterized by a second variation of a suitably defined entropy
functional. As discussed in Padmanabhan (1990), for all isothermal spheres extending up to a
critical radius $R$ so that $E$ is the Newtonian total energy at $R$, the constraint
$R\,E/(G\,M^2)>-0.335$ must hold, though stable configuration with local maxima must
also satisfy $\rho_c<709\,\rho(R)$, while isothermal spheres complying with $R\,E/(G\,M^2)>-0.335$ 
but for which $\rho_c/\rho(R)  > 709$ correspond to metastable configurations in which the entropy extrema is a
saddle point. The critical radius $R$ complies with
\ba \frac{\rho_c}{\rho(R)} \ = \ 709,\label{709}\ea
corresponding to the maximal stable isothermal sphere for which entropy is a local
maximum. Using equation (\ref{lambda}), we can find the ratio
\ba \frac{\rho(R)}{\rhocr} \ = \ \frac{5.55\times
10^3}{h^2}\,\frac{\rho_c}{\textrm{M}_\odot/\textrm{pc}^3},\label{R2cr}\ea
which can be compared to the ratio of $\la\rho\ra (r_{\textrm{\tiny{vir}}})$ to $\rhocr$
given by equation (\ref{rvir0}). This yields approximately
\ba \frac{\la\rho\ra (r_{\textrm{\tiny{vir}}})}{\rho(R)} \ \sim \
\frac{\rho(r_{\textrm{\tiny{vir}}})}{\rho(R)} \ \sim \
\frac{h^2\,\Delta}{5.55\times
10^3}\,\frac{\textrm{M}_\odot/\textrm{pc}^3}{\rho_c},\label{R2vir}\ea
but, if $R< r_1$, then we should have approximately $\la\rho\ra\sim\rho\sim r^{-2}$
for $r<R$, hence we obtain a ratio
\ba \frac{R}{r_{\textrm{\tiny{vir}}}} \ \sim \
\frac{h\,\Delta^{1/2}}{74.16}\,
\left(\frac{\textrm{M}_\odot/\textrm{pc}^3}{\rho_c}\right)^{1/2},\label{Rvir}\ea
and also the following ratio analogous to (\ref{rlv})
\ba \frac{r_1}{R} \ \sim \ \frac{74.16}{(2\,\Omega_\Lambda)^{1/2}\,h}\,
\left(\frac{\rho_c}{\textrm{M}_\odot/\textrm{pc}^3}\right)^{1/2}.\label{Rlambda}\ea
Assuming that $h=0.65,\,\Omega_\Lambda=0.7$, equations (\ref{Rvir}) and (\ref{Rlambda})
show that for values (consistent with current estimates) around $\rho_c\sim
10^{-3}-10^{-2}\, \textrm{M}_\odot/\textrm{pc}^3$ (see figures 5 and 6), the virialization
radius is about the same order of magnitude as the length scale $R$ of equation (\ref{709}) and about one to
two orders of magnitude smaller than $r_{1}$. An estimate of the value of
$R$ in Mpc follows from the fact that the variables (\ref{adimvars}) that we use coincide
(save notation differences, a minus sign in
$\Psi$) with those defined in equations (4.21) and (4.22) of Padmanabhan (1990) which yield
$x(R)=R/r_0=34.2$. Using equation (\ref{r0vals}) we obtain
\ba r(R) \ = \ 1.467 \times 10^{-4} \, \frac{\sigma_c}{\textrm{km/sec}}\,
\left(\frac{\textrm{M}_\odot/\textrm{pc}^3}{\rho_c}\right)^{1/2}\textrm{Mpc}.\nonumber\\
\label{Rvals}\ea
By looking at the numerical solutions for the equilibrium equations
(\ref{feqnewt3}) and (\ref{feqnewt4}), we will verify in section IX the qualitative
estimates of $r_1$,\,$r_2$,\,$r_{\textrm{\tiny{vir}}}$ and $R$ that we
have presented above. These estimates tend to support the conclusion that
the effects of a $\Lambda$ field can be ignored in the study of virialized structures.

\section{The King models}

It is interesting to compare the effect of a $\Lambda$--field on a MB distribution with
other distribution functions (but without $\Lambda$). We examine briefly the case of
the Michie--King models, obtained by truncation of the MB distribution
at a given maximal escape velocity (Binney \& Tremaine 1987). This distribution functions leads to
equilibrium equations similar to (\ref{feqnewt2}), but with $\rho/\rho_c$ given
by Binney \& Tremaine (1987) and Katz (1980)
\ba \frac{\rho}{\rho_c} && = \
\frac{K(\tilde\Psi)}{K(\Psi_0)},\\
\tilde\Psi && \equiv \ \Psi+\Psi_0,\qquad
\Psi_0 \ \equiv \
\frac{\Phi_b-\Phi_c}{\sigma_c^2},\\ K(X)
&& = \
\textrm{e}\,^{X}\,\textrm{erf}(X^{1/2})-
\frac{2\,X^{1/2}}{\sqrt\pi}\,\left(1+\frac{2}{3}
X\right),\ea
where $\Psi$ is given by equation (\ref{defPsi}), while $\Phi_b\equiv \Phi(x_b)$ for $x=x_b$
satisfying the constraint $K(x_b)=0\,\Rightarrow\, \rho(x_b)=0$. The $x_b$ is known as
the ``tidal radius'' marking a theoretical cut off scale whose value depends on $\Psi_0$. The
equilibrium equations for the King models expressed in terms of the dimensionless variables
$x$ and $\M$ are then

\begin{figure}
\psfig{file=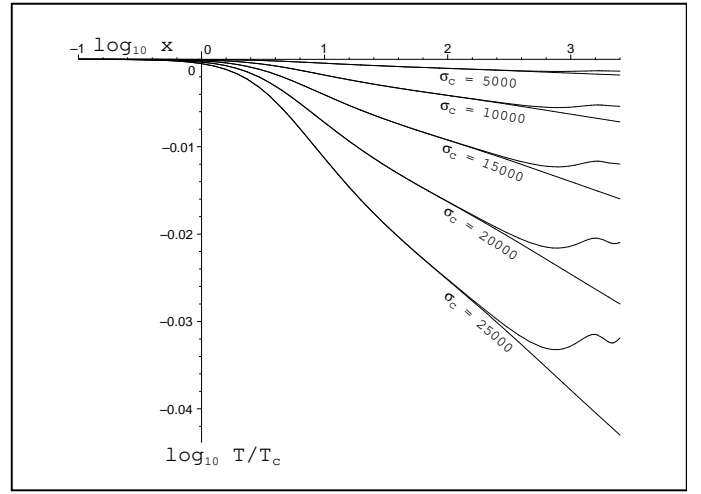,angle=0,width=9.0cm}
\caption{Each pair of curves provides the radial temperature variations within a MB
sphere characterized by a given central velocity dispersion (values indicated in  
$\hbox{km}/\hbox{sec}$). The curves showing monotonous decay correspond to
$\lambda=0$. In the case with nonzero cosmological
$\lambda$ term, the temperature curves stop decreasing and oscilate around a constant temperature 
(see text). It is evident that as the velocity dispersion of the halo decreases,
the halo tends to an isothermal configuration, valid to a high degree in any astrophysical 
structure, $\sigma_{c}<1000\, \hbox{km}/\hbox{sec}$.}
\label{fig1}
\end{figure}

\ba \frac{\dd\M}{\textrm{d}x} &&  = \
x^2\,\frac{K(\tilde\Psi)}{K(\Psi_0)},
\nonumber\\
\frac{\dd\Psi}{\textrm{d}x} && = \
-\frac{\M}{x^2}. \label{feqK}\ea
Since $\tilde\Psi(0)=\Psi_0$ and $\M(0)=0$, the integration of these equations
requires one to specify a given value of $\Psi_0$, leading to a family of
numerical solutions parametrized by $\Psi_0$. Theoretical studies of the King
distribution applied to globular clusters (see Katz 1980) yield the stability range
$\Psi_0 < 8$, though there are observations and theoretical estimates allowing
for values up to $\Psi_0\approx 11$.

As we will show in the following section, the dynamical variables of the
King models behave asymptotically in a very different manner from those of the MB models with
$\Lambda$. However, the cut off scale $r_\lambda$ in the latter is of the same order of
magnitude as the tidal radius of King models with deep potential wells ($\Psi_0\sim 10$).

\begin{figure}
\psfig{file=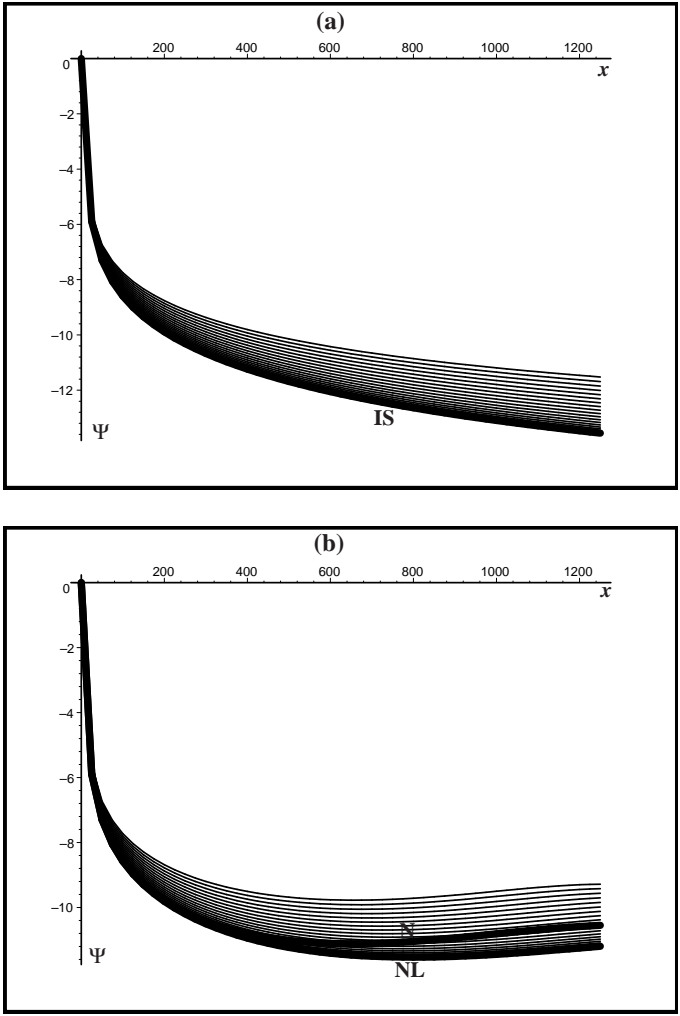,angle=0,width=9.0cm}
\caption{a) Relative potential vs. radius, for MB spheres in the absence of a
$\lambda$ term. The curves are characterized by sequentially decreasing
central velocity dispersions, ranging from relativistic values (upper curves) to newtonian values
(lower curves). The solid curve at the bottom corresponds to the classical isothermal 
sphere solution. As the central velocity dispersion decreases the sequence
accurately converges to the isothermal sphere. b) Same as panel (a), but in the presence of a
cosmological $\lambda$ term. The curves converge to the solid curve marked NL associated with
equations (\ref{feqnewt3}), hence it is clear that this is the correct Newtonian
limit for the MB spheres obtained from the full General Relativity treatment. 
The curve marked N corresponds to the naive Newtonian limit associated with equation (\ref{feqnewt3}).
It is evident that this is not the correct newtonian limit of the MB sphere when there is a 
$\Lambda$--field.}
\label{fig1}
\end{figure}

\begin{figure}

\psfig{file=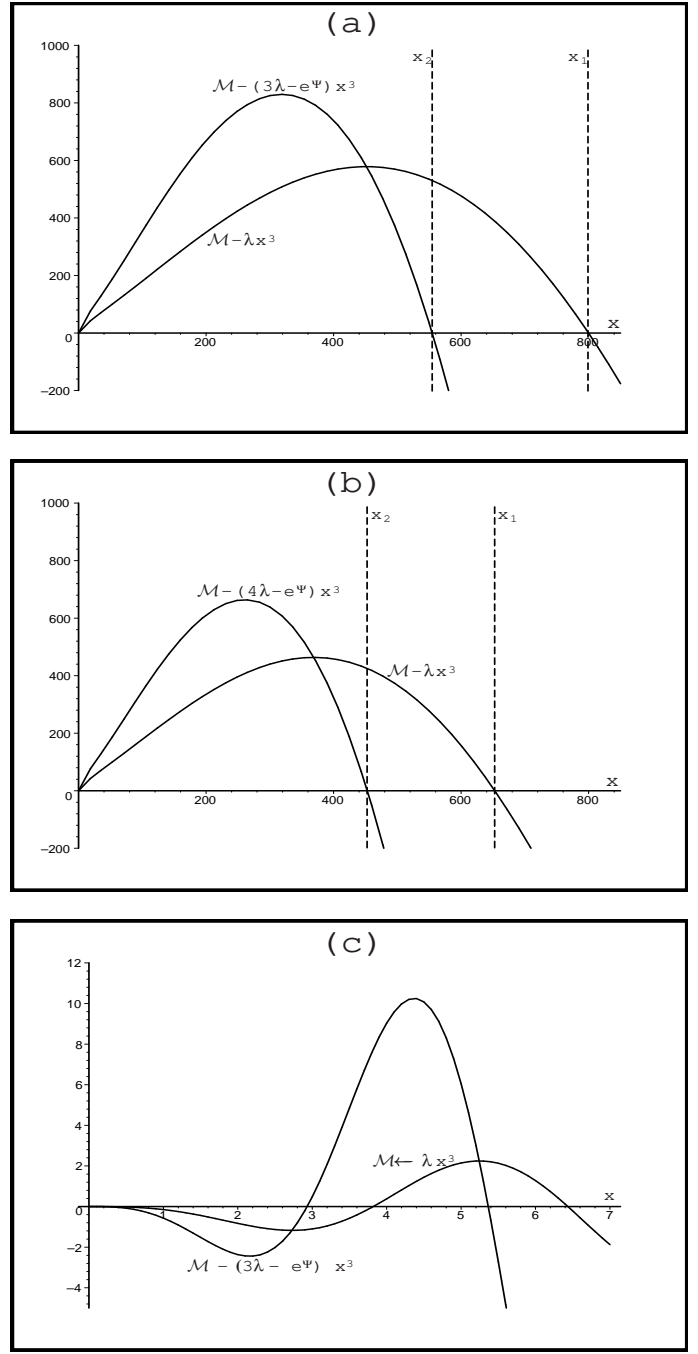,angle=0,width=9.0cm,height=18cm}
\caption{Values of the existence ($M-\lambda x^{3}$) and stability ($M-(3\lambda -e^{\Psi})x^{3}$)
conditions for circular orbits vs. the dimensionless radial coordinate $x$ for MB spheres with
$\rho_{c}=0.01 \hbox{M}_{\odot}/\hbox{pc}^3$. Panels (a) and (b) respectively depict the cases of the 
Newtonian limit and the naive Newtonian treatment. In the latter case the existence and stability 
radii are about two thirds smaller than those of (a). Panel (c) For depicts the case where
central density is lower than the threshold $\rho_{c}>2 \Omega_{\lambda}$. In this case no central
structure can exist, as the existance and stability conditions are only met in disconnected ranges
not including the origin.}
\label{fig1}
\end{figure}

\section{Numerical analysis}

We solve numerically the systems of equations (\ref{feqrel}), (\ref{feqnewt2}), (\ref{feqnewt3}),
(\ref{feqnewt4}) and (\ref{feqK}), assuming the regularity conditions
$\M_c=\M'_c=\Psi_c=\Psi'_c=0$, for $\Omega_\Lambda=0.7,\,h=0.65$ and suitable values of the
free parameters $\sigma_c,\,\rho_c,\,\Psi_0$.

\subsection{The isothermal limit with a $\Lambda$ field}

It is interesting to examine the Newtonian limit of the MB gas (with and without the
$\Lambda$--field) by looking at how an isothermal behavior emerges as temperature gradients
decrease for sequentially decreasing values of the ratio $\sigma_c/c$, starting from
relativistic values down to smaller Newtonian velocities. Using $\Psi$ from the numerical
solution of equation (\ref{feqrel}), we plot in figure 1 the normalized temperature, $T/T_c$,
given by equation (\ref{temp}) vs
$\log_{10}\,x$  for four values in the range $\sigma_c/c= 0.08$ down to
$\sigma_c/c=1.6\times 10^{-2}$, that is from $\sigma_c=25000$ km/sec to $\sigma_c=5000$
km/sec (still a dispersion velocity much larger than that of any galactic structure). As
shown in this figure, the temperature gradients of the MB gas are already small
($T/T_c < 0.2$) even for $\sigma_c = 25000$ km/sec, suggesting that even for
more relativistic values of $\sigma_c/c$ the linearized expression (\ref{tempNL}) can be
used instead of equation (\ref{temp}). These gradients steadily decrease for smaller
$\sigma_c/c$, becoming negligible for the velocities characteristic of galactic
structures, and so approaching in the limit $\sigma_c/c\to 0$ an isothermal condition
$T=T_c$. In the case $\Lambda=0$ the temperature curves steadily decrease
so that $T/T_c\to 0$ as $x\to\infty$. However, if there is a $\Lambda$--field, then
these curves stop decreasing at $x=x_1$ and begin to oscillate around a constant
value, $T=T_\lambda$, that can be identified as a characteristic
temperature of the $\Lambda$--field.

\subsection{The correct Newtonian limit}

The Newtonian limit of equation (\ref{feqrel}) can also be illustrated by means of sequences of
curves corresponding to $\Psi$ obtained from equation (\ref{feqrel}) for sequentially decreasing
values of $\sigma_c/c$. The curves of figure 2a ($\lambda=0$) clearly tend to the thick
curve at the bottom, marked as ``IS'', corresponding to $\Psi$ for the isothermal sphere,
obtained from equation (\ref{feqnewt2}). In figure 2b we consider the case with nonzero
$\lambda$ given by equation (\ref{lambda}). It is evident that the sequence of curves tends to
the thick curve, marked NL, corresponding to $\Psi$ obtained from equation
(\ref{feqnewt3}) and not to the thick curve, marked ``N'', that corresponds to $\Psi$
obtained from equation (\ref{feqnewt4}). Hence, the correct Newtonian limit of a MB gas with a
$\Lambda$ field is given by equations (\ref{feqnewt3})--(\ref{poissonNL}) and not by the Newtonian
equations (\ref{feqnewt4})--(\ref{poissonPN}).

\subsection{The ``cut off'' length scales}

Numerical solution of the systems (\ref{feqnewt3}) and (\ref{feqnewt4}) allow us to
depict graphically conditions (\ref{condexist})--(\ref{condstab}) (figures 3a and 3b),
conditions (\ref{condexist_ave})--(\ref{condstab_ave}) giving the same results, where the length scales for
the existence and stability of circular orbits, $x_1$ and $x_2$, follow from the roots given by
equations (\ref{xlambda_1})--(\ref{xlambda_2}) and (\ref{xlambda1_ave})--(\ref{xlambda2_ave}).
Assuming $\rho_c=0.01\,\textrm{M}_\odot/\textrm{pc}^3$, the following numerical values emerge
from figure 3a: \, $x_1\approx 800$, \, $x_2\approx 555$, \, while for the Newtonian
equations (\ref{feqnewt4}) we obtain smaller values (see figure 3b): $x_1\approx 660$, \, $x_2\approx
460$. For the same value of $\rho_c$ and for the Newtonian limit equations (\ref{feqnewt3}) we
obtain from figure 5: $x_{\textrm{\tiny{vir}}}\approx 72$ and $x(R)\approx 35$. These values of $x$
can be translated into actual distances for specific galactic structures by using $r=r_0\,x$ where
$r_0$ is given by equation (\ref{r0vals}). The resulting length scales are shown in Table 1.

\begin{figure}
\psfig{file=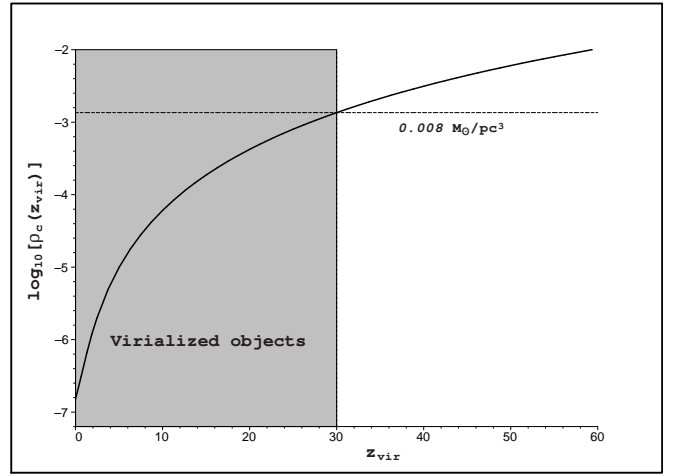,angle=0,width=9.0cm}
\caption{The solid curve shows the minimum central density for stability as a function of
virialization redshift. The value at $z=30$, corresponding to $\rho_{c}=0.01 M_{\odot} pc^{-2}$, is
of the order of observed values in galactic structures which can be thought of as having
formed at around that time or less.}
\label{fig1}
\end{figure}

\begin{figure}
\psfig{file=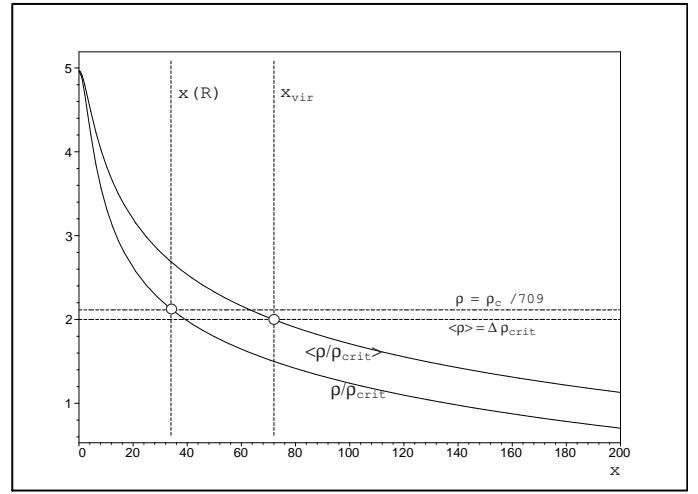,angle=0,width=9.0cm}
\caption{Logarithmic plots of the density and average density 
vs. dimensionless radius $x$, both normalized by $\rhocr$. The intersects with the
horizontal dotted lines define the onset of Antonov instability ($\rho=\rho_{c}/709$) and the virial
radius ($<\rho>=\Delta \rho_{crit}$).}
\label{fig1}
\end{figure}

As we mentioned in section VI, the conditions (\ref{condexist})--(\ref{condstab}) for the existence
and stability of circular geodesic orbits are violated for $\rho_c < 2\,\Omega_\Lambda\,\rhocr $.
This possibility is illustrated by figure 3c, where these conditions are plotted for $\rho_c =
\Omega_\Lambda\,\rhocr$, clearly showing how conditions (\ref{condexist})--(\ref{condstab}) do not hold
around the center (though they do hold in disconnected ranges of $x$ that exclude the center).
Since both $\rho_c$ and $\rhocr$ are functions of the cosmic era, condition (\ref{min_lam}) with
the help of equation (\ref{rhocrz}) yields a minimal bound of $\rho_c$ for virialized structures complying
with conditions (\ref{condexist})--(\ref{condstab}). This is illustrated in figure 4, showing that structures
of relatively recent virialization easily meet this minimal bound, though for galaxies which virialized at
$z\sim 30$, the fulfillment of condition (\ref{min_lam}) is non--trivial, leading to $\rho_c > 0.008\,
\,\textrm{M}_\odot/\textrm{pc}^3$, a value that is consistent with current estimates of $\rho_c$.

In figures 3a--b and 5 we considered $\rho_c= 0.01\,\,\textrm{M}_\odot/\textrm{pc}^3$. In order
to find out the dependence of the values of $x_1,\,x_2$ and $R$ on $\rho_c$, we plot in figure
6 the ratios of these length scales with respect to $x_{\textrm{\tiny{vir}}}$, clearly
emerging from this figure that $x_1$ and $x_2$ have a very weak dependence on $\rho_c$ and are,
respectively, about 10 and 7 times larger than $x_{\textrm{\tiny{vir}}}$ (for the Newtonian case
these figures are about two thirds smaller). However, $R$ is quite sensitive to $\rho_c$, with
$x_{\textrm{\tiny{vir}}} \sim R$ for densities in the range $0.001-0.01\,\,
\textrm{M}_\odot/\textrm{pc}^3$. This is a nice result, illustrating that these two physically
motivated length scales approximately coincide for values of $\rho_c$ close to current estimates.
All length scales we have considered show a clear linear dependence on $\sigma_c$ (from equation
(\ref{r0vals})). This is shown in figure 7, displaying the graph of each length scale
as a function of $\sigma_c$ for a wide range of values of $\rho_c$
($10^{-5}-1\,\,\textrm{M}_\odot/\textrm{pc}^3$). Since the dependence on $\rho_c$ is very weak, the
strip of lines for all values of $\rho_c$ simply makes the lines thicker.

\begin{table}
\begin{center}
\begin{tabular}{|c|c|c|c|}
\hline
{Length scale} & {Dwarf } & {Large } & {Cluster} \\
{} & {galaxy} & {galaxy} & {} \\ \hline\hline
{Existence of circular\hfill} & {} & {} & {} \\
{orbits, $r_1$\hfill} & {140 kpc} & {5 Mpc} & {32 Mpc}\\ \hline
{Largest stable circular\hfill} & {} & {} & {}\\
{orbit, $r_2$\hfill} & {90 kpc} & {3 Mpc} & {22 Mpc}\\ \hline
{Virial radius, $r_{\textrm{\tiny{vir}}}$\hfill} & {} & {} & {}\\
{} & {14 kpc} & {300 kpc} & {3 Mpc}\\ \hline
{Antonov instability, $R$\hfill} & {} & {} & {}\\
{ $\rho_c = 0.01\,\textrm{M}_\odot/\textrm{pc}^3$\hfill} & {8 kpc} & {140 kpc} & {1.3 Mpc}\\ \hline
{Antonov instability, $R$\hfill} & {} & {} & {}\\
{$\rho_c = 0.001\,\textrm{M}_\odot/\textrm{pc}^3$\hfill} & {15 kpc} & {330 kpc} & {3.2 Mpc}\\ \hline
\end{tabular}
\end{center}
\caption{{\em Length scales for galactic structures. Dwarf and large galaxies and clusters are
respectively characterized by $\sigma_c=3,\,200,\,1000$ km/sec. The length scales $r_1,\,r_2$ and
$r_{\textrm{\tiny{vir}}}$ were evaluated for $\rho_c = 0.01\,\textrm{M}_\odot/\textrm{pc}^3$. See
also figures 6 and 7.}}
\end{table}

\subsection{Density profiles and rotation curves}

The effects of the $\Lambda$--field can be clearly illustrated by plotting relevant
physical variables of the isothermal halos, such as the density profiles and rotation velocity along circular
orbits.

\begin{figure}
\psfig{file=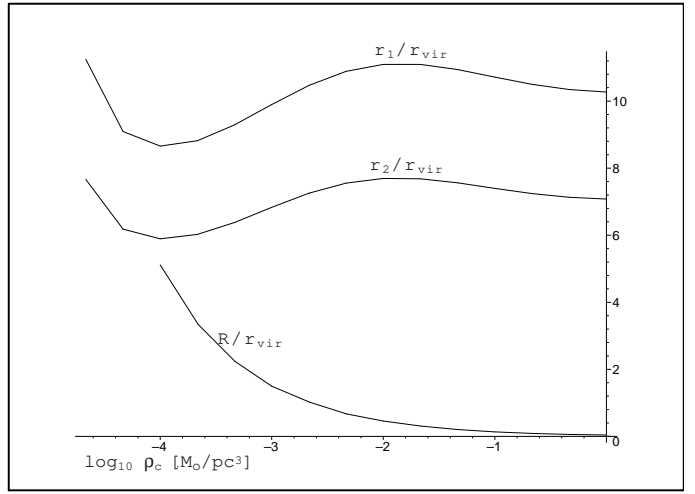,angle=0,width=9.0cm}
\caption{Dependence of circular orbits existence and stability radii, $r_{1}$ and $r_{2}$, and onset
of Antonov instability, $R$, all in units of the virial radius, as functions of the logarithm of 
central density. It is evident that both $r_{1}$ and $r_{2}$ depend only very weakly on 
$\rho_c$ and are always well beyond the virial radius. The radius $R$ is much more sensitive to 
central density, being rather small for larger $\rho_c$. Since a heat conduction mechanism 
(interaction cross-section) is required for this mechanism 
to operate, it probably does not play a significant role.}
\label{fig1}
\end{figure}

Figure 8 depicts a logarithmic plot of the rest--mass density, $\rho$, normalized by
$\rhocr$ for $\rho_c =0.01 \,\textrm{M}_\odot/pc^3$, in comparison with the rest-mass density of the
isothermal sphere and of King halos associated with various values of $\Psi_0$ (only the Newtonian
limit ``NL'' case is shown). It is evident that in the range $0<x<x_2$ where stable circular orbits
exist, the density curve of the NL case is almost identical with that of the isothermal sphere
(marked as ``IS''). However, in the asymptotic range $x>x_1$, these curves are completely different,
since for the IS
$\rho$ simply decays as $x^{-2}$ while in the NL case it oscillates around a value close to
$\rhocr$. Since $\rho=\rho_c\,\textrm{e}^\Psi $, these oscillations mark relative maxima
and minima of the Newtonian potential, hence $x=x_1$ marks an unstable equilibrium where
this potential becomes repulsive (though furhter out it alternates from repulsive to attractive).
Clearly, this behavior of the asymptotic region is dominated by the  $\Lambda$--field and
is in stark contrast with the behavior of the isothermal region $0<x<x_2$. The region
$x_2<x<x_1$, where circular orbits exist but are unstable, is a transition zone between the
isothermal and asymptotic zones. However, in the physical range $0<x< x_{\textrm{\tiny{vir}}}$, the
effects of the $\Lambda$--field are practically negligible. Notice that
$\rho$ for the NL case is also very different from the density curves of the King halos, though the
cut off boundary (the tidal radius) of some of these halos ($\Psi_0 > 10$) can be of the same
order of magnitude (or even larger) than $x_2$.

\begin{figure}
\psfig{file=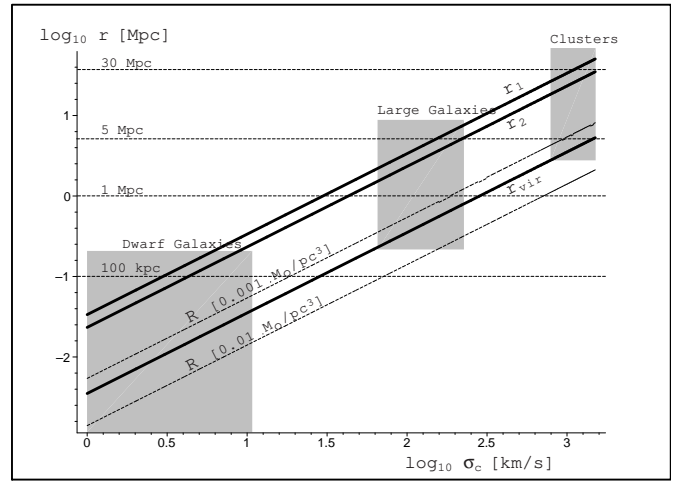,angle=0,width=9.0cm}
\caption{The plot depicts the relation between the indicated radii vs. velocity dispersion, 
where we have
marked with shaded regions the zones occupied by various astrophysical structures. The solid lines 
represent existence, stability
and virial radii for a wide range of central densities. The dotted lines correspond to radii for the
onset of Antonov instability for the two indicated values of the central density.}
\label{fig1}
\end{figure}

\begin{figure}
\psfig{file=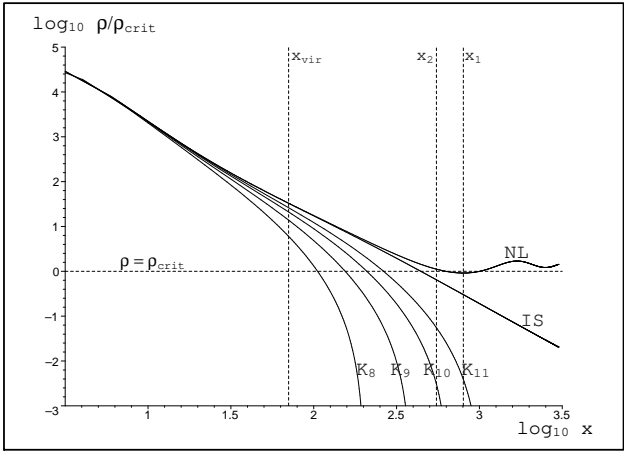,angle=0,width=9.0cm}
\caption{Density profiles normalized by $\rhocr$ as functions of the dimensionless radius $x$. 
Curves labeled by {\bf K8} to {\bf K11} give classical King profiles of the indicated shape parameter, 
{\bf IS} is the  isothermal sphere, and {\bf NL} corresponds to the newtonian limit of an MB gas
in the presence of a cosmological $\lambda$ term. Vertical dotted
lines indicate the virial radius and the radii for stability and existence of circular orbits.}
\label{fig1}
\end{figure}

Rotation velocities normalized by $\sigma_c$ are given by equation (\ref{rotvel}). These ``rotation curves''
are plotted in figure 9 for the NL case for $\rho_c =0.01\,\textrm{M}_\odot/pc^3$, in comparison
with rotation curves of the isothermal sphere and of King halos associated with various values of
$\Psi_0$. The figure also displays the cut off length scales $x_1,\,x_2$, separating the isothermal,
asymptotic and transition zones, as well as the physical radius $x_{\textrm{\tiny{vir}}}$. Since
this is not a logarithmic plot, the effects of the $\Lambda$--field are more noticeable. For example,
in the NL case the rotation curve is not flat and decreases significantly already in the isothermal
region, though it is practically identical to the rotation curve of the isothermal sphere in the
range $0<x< x_{\textrm{\tiny{vir}}}$ where all visible matter would be located. The rotation curve of
the NL case plunges to zero for $x=x_1$, so that for a range of values $x>x_1$ circular geodesic
orbits do not exist. This is connected to the fact that $x=x_1$ marks an unstable equilibrium where
the Newtonian potential becomes repulsive. However, this potential oscillates, hence it becomes
again attractive and then repulsive again, all of this in disconnected domains of
$x$ for $x>x_1$. As a comparison, the rotation curves of King halos (evaluated up to their tidal
radii) also decrease but never reach zero.

\begin{figure}
\psfig{file=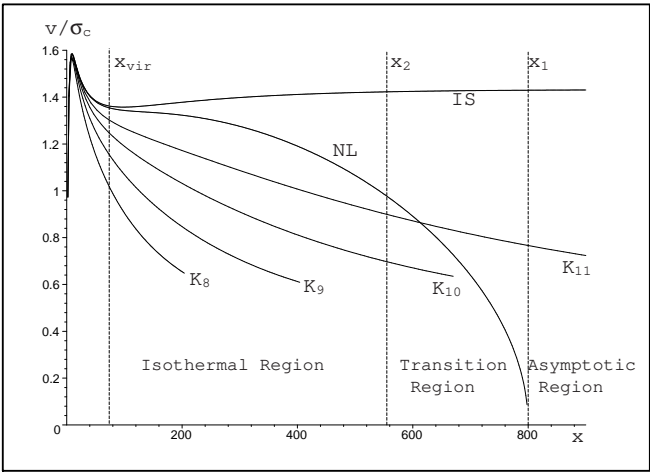,angle=0,width=9.0cm}
\caption{Rotation velocity curves for the cases shown in figure 8. The graphs illustrate that within
the Isothermal region, {\bf NL} follows {\bf IS} to within a factor of 1.4, however, within the 
Transition region, the {\bf NL} rotation curve falls to zero. (see text)}
\label{fig1}
\end{figure}

\section{Conclusion}

We have examined the equilibrium of a MB gas in the presence of a
$\Lambda$--field constrained by recent observational
data. Proceeding from a general relativistic framework and carefully taking the
Newtonian limit, we obtain the equilibrium equations (\ref{feqnewt3})--(\ref{poissonNL}) that are
different form those that follow from a ``naive'' Newtonian treatment that would simply add the
$\Lambda$--field to an isothermal sphere (i.e. equations (\ref{feqnewt4})--(\ref{poissonPN})).
We have shown that the presence of this repulsive field implies that isothermal
configurations are now characterized by  the length scales, $x_1,\,x_2$, defined by conditions
(\ref{condexist}) and (\ref{condstab}) and respectively associated with existence and stability of
circular geodesic orbits. We have shown that these length scales can also be characterized by
the constraints given by (\ref{condexist_ave}) and (\ref{condstab_ave}) on the rest--mass density and
its volume average, hence $x_2\leq x_1$. The fulfillment of the existence
and stability constraints in a region containing the symmetry center $x=0$ yields a minimal central
rest--mass density $\rho_c=2\,\Omega_\Lambda\,\rhocr$, and thus a minimal central density of $\rho_c
= 0.008 \textrm{M}_\odot/\textrm{pc}^3$ for galactic structures having virialized at $z\sim 30$.
Since $x=x_2$ marks the radius of the largest stable circular orbit, it is reasonable to consider
this radius as the cut off length scale of isothermal configurations in equilibrium with a
$\Lambda$--field.  The numerical solution of the equilibrium equations provides precise
numerical estimates of all involved length scales, showing that $x_2$ is about 7 times larger than
physically motivated cut off length scales of isothermal halos, such as the virialization radius
$x_{\textrm{\tiny{vir}}}$ and the radius for the onset of Antonov instability $R$. By looking at
the graphs of the rest--mass density and rotation curves, we can
identify isothermal ($0<x<x_2$), asymptotic ($x>x_1$) and transition ($x_2<x<x_1$) regions in which
these variables change from an almost isothermal behavior to an asymptotic regime dominated by the
$\Lambda$--field. It is quite noticeable that up to the physical cut off scale,
$x_{\textrm{\tiny{vir}}}$, the behavior of the MB gas with the
$\Lambda$--field is practically indistinguishable from that of an isothermal sphere.
Therefore, for the purposes of studying the equilibrium state of virialized structures, the effects
of the $\Lambda$--field are negligible and can be safely ignored. However, these effects might be
important in the study of the gravitational clustering of larger structures that have already left
the linear regime but are not yet virialized, or the formation of galactic structures (though
the assumption of thermodynamical equilibrium might not be applicable in these cases).

Finally, the zero gradient approximation associated with the constant
$\Lambda$--field is basically a toy model of more general quintessence--like sources
described by non--trivial scalar fields. The equilibrium of a MB gas together with
such a source could yield interesting features that are absent in the
$\Lambda$--field approximation. This problem is currently under investigation (N\'u\~ez, Matos and 
Sussman, 2003).

\section{ACKNOWLEDGMENTS}

XH acknowledges financial support from {\bf CoNaCyT} grant {\tt  
I39181-E}.
RAS is thankful to Offer Lahav for his warm hospitality and  
illuminating discussions. RAS also acknowledges inspiration from the  
wise meowing of his feline friends Moquis, Tontis and Ni\~na.

\end{document}